\documentclass[superscriptaddress,nofootinbib,showpacs,showkeys,longbibliography,hyperref,notitlepage]{revtex4-1}

\usepackage[margin=1in]{geometry}
\usepackage{graphicx}
\usepackage{graphicx,color}
\usepackage{amsmath,amssymb,amsfonts}
\usepackage[plainpages=false]{hyperref}

\def\Abstract#1{\noindent {\normalsize {\bf Abstract:} {\normalfont #1}}}
\def\Conference{\vspace{4mm}\begin{raggedright} {\normalsize {\it Talk presented at the 2019 Meeting of the Division of Particles and Fields of the American Physical Society (DPF2019), July 29--August 2, 2019, Northeastern University, Boston, C1907293.} } \end{raggedright}\vspace{4mm}}

\def \be {\begin{equation}}
\def \ee {\end{equation}}
\def \ee  {\end{equation}}
\def \bea {\begin{eqnarray}}
\def \eea {\end{eqnarray}}

\newcommand{\pT} {\ensuremath{p_{\mathrm{T}}}}

\begin{document}

%%%%%%%%%%%%%%%%%%%%%%%%%%%%%%%%%%%%%%%%%%%%%%%%%%%%%%%%%%%%%%%%%%%%%%%%%%%
%
% TITLE, AUTHOR, INSTITUTION, ABSTRACT ==> UPDATE
% 
%%%%%%%%%%%%%%%%%%%%%%%%%%%%%%%%%%%%%%%%%%%%%%%%%%%%%%%%%%%%%%%%%%%%%%%%%%%

\title{Beam-energy dependence of the azimuthal anisotropic flow from RHIC}
\medskip

\author{Niseem~Magdy~(For the STAR Collaboration)} 
\email{niseemm@gmail.com}
\affiliation{Department of Physics, University of Illinois at Chicago, Chicago, Illinois 60607, USA}

\maketitle

\Abstract{
Recent STAR measurements of the anisotropic flow coefficients (v$_{n}$) are presented for Au+Au collisions spanning the beam energy range $\sqrt{s_{NN}} = 7.7 - 200$~GeV. The measurements indicate dependences on the harmonic number ($n$), transverse momentum ($p_T$), pseudorapidity ($\eta$), collision centrality and beam energy ($\sqrt{s_{NN}}$) which could serve as important constraints to test different initial-state models and to aid precision extraction of the temperature dependence of the specific shear viscosity.
}

\Conference

%%%%%%%%%%%%%%%%%%%%%%%%%%%%%%%%%%%%%%%%%%%%%%%%%%%%%%%%%%%%%%%%%%%%%%%%%%%
%
% MAIN TEXT ==> UPDATE
% 
%%%%%%%%%%%%%%%%%%%%%%%%%%%%%%%%%%%%%%%%%%%%%%%%%%%%%%%%%%%%%%%%%%%%%%%%%%%

\section{Introduction}

A major aim of the heavy-ion experimental program at the Relativistic Heavy Ion Collider (RHIC) is to study the properties of the quark-gluon plasma (QGP) created in ion-ion collisions. Recently, several studies have highlighted the use of anisotropic flow measurements to investigate the transport properties of the QGP ~\cite{Teaney:2003kp,Lacey:2006pn,Schenke:2011tv, Song:2011qa,Niemi:2012ry,Qin:2010pf,Magdy:2017kji}. An essential question in many of these studies has been the role of initial-state fluctuations and their impact on the uncertainties associated with the extraction of $\eta/s$ for the QGP~\cite{Alver:2010gr,Lacey:2013eia}. 

In this work, we present a new measurement for the anisotropic flow coefficients, v$_{n}$($n > 1$)~\cite{Magdy:2018itt,Adam:2019woz,Magdy:2018ufy}, and the rapidity-even dipolar flow coefficient, v$^{even}_{1}$~\cite{Star:2018zpt,Magdy:2018whk}, with an eye toward providing a new constraint which could assist the distinction between different initial-state models and therefore, facilitate a more accurate extraction of the specific shear viscosity, $\eta/s$~\cite{Auvinen:2017fjw,Auvinen:2017pny}. 

The anisotropic flow is described by the coefficients,  v$_{n}$, obtained from a Fourier expansion of the azimuthal angle ($\phi$) distribution of the particles emitted in the collisions~\cite{Poskanzer:1998yz}:

\begin{eqnarray}
\label{eq:1}
\frac{dN}{d\phi}\propto1+2\sum_{n=1}\mathrm{v_{n}}\cos (n(\phi-\Psi_{n})),
\end{eqnarray}
where $\Psi_n$ denotes the azimuthal angle of the $n^{th}$-order event plane; the coefficients, v$_{1}$, v$_{2}$ and v$_{3}$ define directed,  elliptic, and triangular flow, respectively. The flow coefficients,  v$_{n}$, are linked to the two-particle Fourier coefficients, v$_{n,n}$, as:

\begin{eqnarray}
\label{eq:3}
\mathrm{v_{n,n}}(\pT^{a},\pT^{b})  = \mathrm{v_n}(\pT^{a})\mathrm{v_n}(\pT^{b})+ \mathrm{\delta_{NF}},
\end{eqnarray}
where a and b are particles with $\pT^{a}$ and $\pT^{b}$, respectively, and $\mathrm{\delta_{NF}}$ is the  non-flow (NF) term, which  involves potential short-range contributions from resonance decays, Bose-Einstein correlation,  near-side jets, and long-range contributions from the global momentum conservation (GMC) ~\cite{Lacey:2005qq,Borghini:2000cm,ATLAS:2012at}. 
The short-range non-flow contributions can be reduced by applying a pseudorapidity gap, $\Delta\eta$, between $\eta^{a}$ and $\eta^{b}$. However, the impacts of the GMC must be explicitly considered. For the current analysis, a simultaneous fitting method~\cite{Star:2018zpt}, outlined below, was used to account for the GMC.

%%%%%%%%%%%%%%%%%%%%%%%%%%%%%%%%%%%%%%%%%%%%%%%
%%%%%%%%%%%%%%%%%%%%%%%%%%%%%%%%%%%%%%%%%%%%%%%
\section{Measurements}
\label{Measurements}
%%%%%%%%%%%%%%
The correlation function method was used to measure the two-particle $\Delta\phi$ correlations:
\begin{eqnarray}\label{corr_func}
 C_{r}(\Delta\phi, \Delta\eta) = \frac{(dN/d\Delta\phi)_{same}}{(dN/d\Delta\phi)_{mixed}},
\end{eqnarray} 
where  $(dN/d\Delta\phi)_{same}$ denotes the normalized azimuthal distribution of  particle pairs  from the same event and $(dN/d\Delta\phi)_{mixed}$ denotes the normalized azimuthal distribution for particle pairs in which each member  is selected  from a different  events but with a similar classification for the collision vertex location,  centrality, etc. The pseudorapidity gap requirement  $|\Delta\eta| > 0.7$ was applied to track pairs to reduce the non-flow contributions associated with the short-range correlations.

The two-particle Fourier coefficients, v$_{n,n}$, are extracted from the correlation function as:
\begin{eqnarray}\label{vn}
\mathrm{v_{n,n}} &=& \frac{\sum_{\Delta\phi} C_{r}(\Delta\phi, \Delta\eta)\cos(n \Delta\phi)}{\sum_{\Delta\phi}~C_{r}(\Delta\phi, \Delta\eta)},
\end{eqnarray}
and  then the two-particle Fourier coefficients, v$_{n,n}$, are used to extract v$^{even}_{1}$ via a simultaneous fit of v$_{1,1}$ as a function of $p_{T}^{\text {b}}$, for several selections of  $p_{T}^{a}$ with Eq.~\ref{eq:3}:

\begin{eqnarray}\label{corrv1}
\mathrm{v_{1,1}}(\pT^{a},\pT^{b})  &=& \mathrm{v^{even}_{1}}(\pT^{a})\mathrm{v^{even}_{1}}(\pT^{b}) - C\pT^{a}\pT^{b}.
\end{eqnarray}
Here, $C \propto 1/(\langle Mult \rangle \langle p_{T}^{2}\rangle)$ takes into account the non-flow correlations caused by a global momentum conservation~\cite{ATLAS:2012at,Retinskaya:2012ky} and $\langle Mult \rangle$ is the mean multiplicity. For  a particular centrality selection, the left-hand side of  Eq.~\ref{corrv1} describes the $N \times N$ matrix which we fit with the right-hand side using $N + 1$ parameters; N values of v$^{even}_{1}(\pT)$ and one additional parameter $C$, accounting for the momentum conservation~\cite{Jia:2012gu}.  

%%%%%%%%%%%%%%%%%%%%%%%%%%%%%%%%%%%%%%%%%%%%%%%
%%%  Fig-1
%%%%%%%%%%%%%%%%%%%%%%%%%%%%%%%%%%%%%%%%%%%%%%%
\begin{figure*}[tb]
\centering{
\vskip -0.2cm
\includegraphics[width=0.9\linewidth,angle=0]{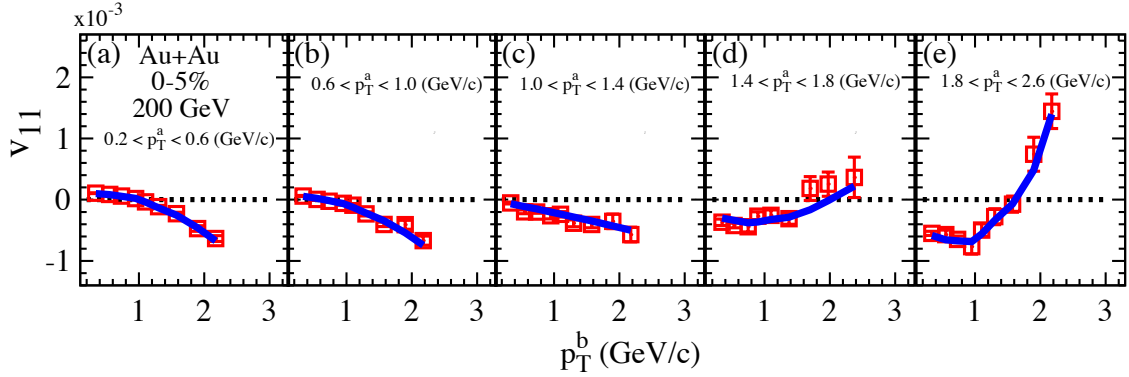}
\vskip -0.3cm
\caption{v$_{1,1}$ vs. $p_{T}^{b}$ for several selections of $p_{T}^{a}$  for 
0-5\% central Au+Au collisions at $\sqrt{s_{_{NN}}} = 200$~GeV. The dashed curves show the result of the simultaneous fit with Eq.~\ref{corrv1}. Figure are taking from Ref~\cite{Star:2018zpt}.
 \label{fig1}
 }
}
\vskip -0.6cm
\end{figure*}

Figure~\ref{fig1}~\cite{Star:2018zpt} shows the result of this fitting method for $0-5\%$ central Au+Au collisions at $\sqrt{s_{_{NN}}} = 200$~GeV. The dashed curve (produced with Eq.~\ref{corrv1}) in each panel represents the effectiveness of the simultaneous fits, as well as the data constraining power. That is, v$_{1,1}(\pT^{b})$ grows from negative to positive values as the selection range for  $\pT^a$ is increased.
%%%%%%%%%%%%%%
\section{Results}
\label{Results}
\begin{figure*}[tb]
\centering{
\vskip -0.2cm
\includegraphics[width=0.90\linewidth,angle=0]{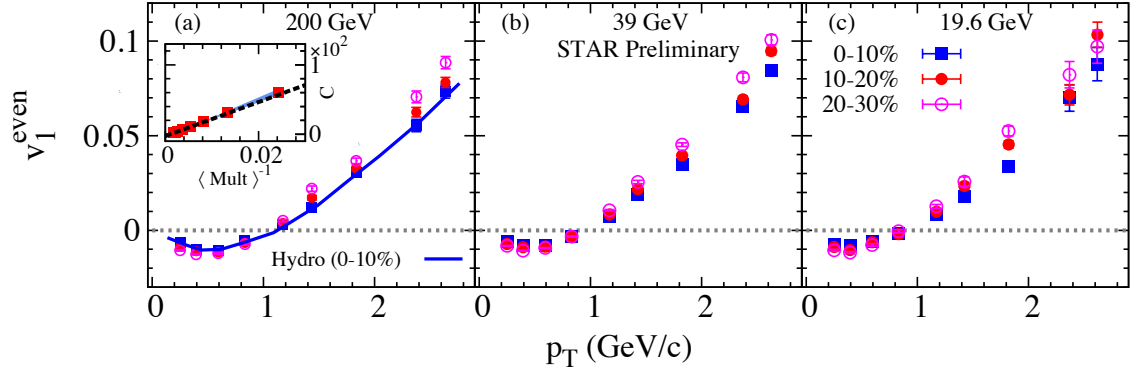}
\vskip -0.2cm
\caption{The extracted values of v$^{even}_{1}$ vs. $p_T$ for different centrality selections (0-10\%, 10-20\% and 20-30\%) of Au+Au collisions for several values of  $\sqrt{s_{_{NN}}}$. The v$^{even}_{1}$ values are obtained via fits with Eq.~(\ref{corrv1}). The solid line in panel (a) shows the result from a hydrodynamic calculations with $\eta/s~=~ 0.16$ \cite{Retinskaya:2012ky}.  The inset in panel (a) shows a representative set of the associated values of $C$ vs. the corrected mean multiplicity for $|\eta| < 0.5$ ($\langle Mult\rangle^{-1}$). The extracted v$^{even}_{1}$ for $\sqrt{s_{_{NN}}}$ = 200, 39 and 19.6 GeV are shown in panels a, b and c respectively.
 \label{fig2}
 }
}
\vskip -0.4cm
\end{figure*}

\begin{figure*}%\vspace{-1.29cm}\hspace{-1.92cm}
\centering{
\vskip -0.2cm
\includegraphics[width=0.90\linewidth,angle=0]{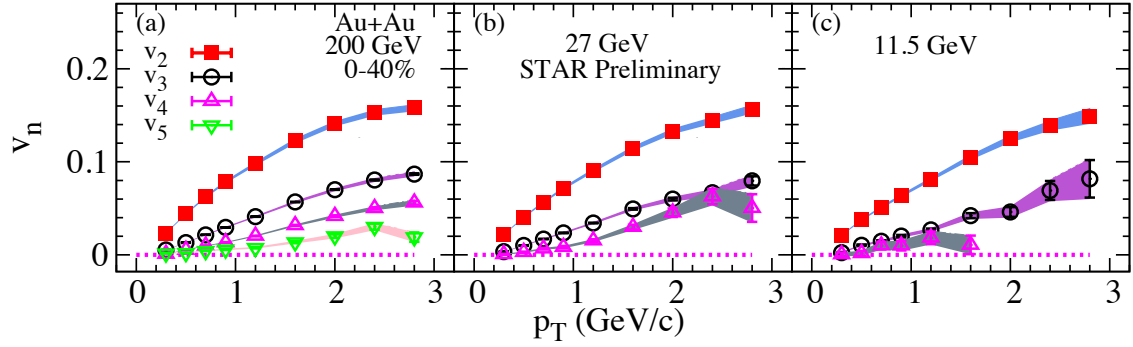}
\vskip -0.2cm
\caption{The v$_{n}(p_{T})$ as a function of $p_{T}$ for charged particles in 0-40\% central Au+Au collisions. The shaded bands represent the systematic uncertainty. The measured  v$_{n}(p_{T})$ for $\sqrt{s_{_{NN}}}$ = 200, 27 and 11.5 GeV are shown in panels a, b and c respectively.
 \label{fig3}
 }
}
\vskip -0.4cm
\end{figure*}

\begin{figure*}%\vspace{-1.29cm}\hspace{-1.92cm}
\centering{
\vskip -0.0cm
\includegraphics[width=0.90\linewidth,angle=0]{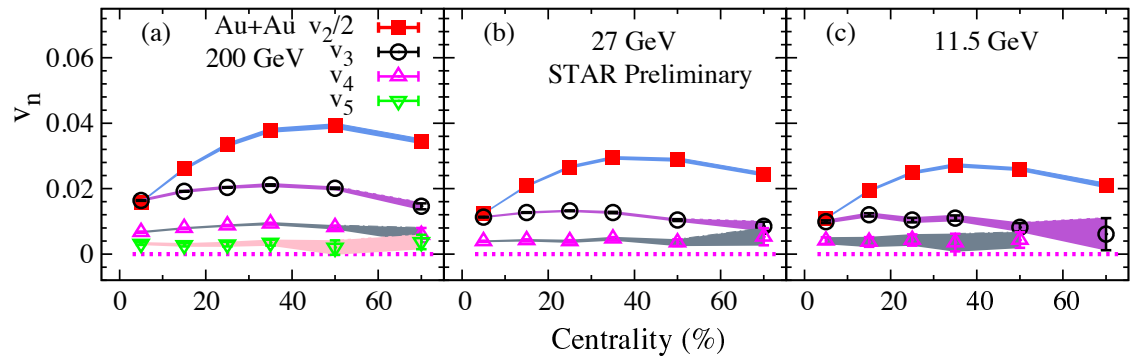}
\vskip -0.2cm
\caption{The  v$_{n}$ as a function of Au+Au collision centrality for charged particles with $0.2 < p_{T} < 4$~GeV/c. The shaded bands represent the systematic uncertainty. The measured  v$_{n}$(Centrality\%) for $\sqrt{s_{_{NN}}}$ = 200, 27 and 11.5 GeV are shown in panels a, b and c respectively.
 \label{fig4}
 }
}
\vskip -0.4cm
\end{figure*}

\begin{figure*}
\centering{
\includegraphics[width=0.4\linewidth,angle=-90]{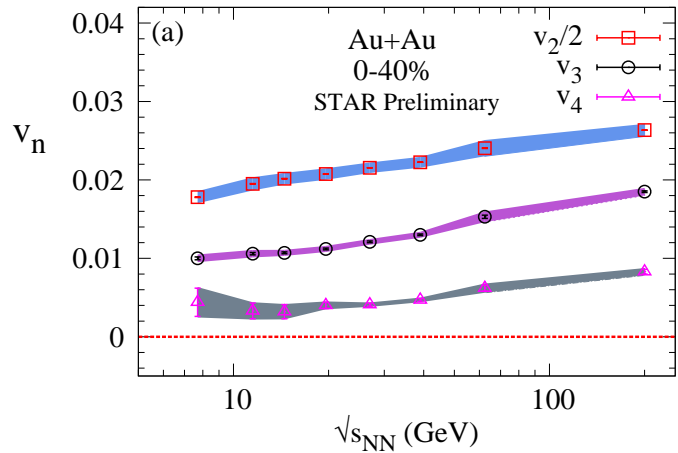}
\vskip -0.2cm
\caption{The v$_{n}(\sqrt{s_{\rm NN}})$ for charged particles with $0.2 < p_{T} < 4$~GeV/c and 0-40\% central Au+Au collisions. The shaded bands represent the systematic uncertainty. The dashed line at v$_{n}$ =0 is to guide the eye.
 \label{fig5}
 }
}
\vskip -0.2cm
\end{figure*}

Representative set of STAR measurements for v$^{even}_{1}$ and v$_{n}$($n \ge 2$) for Au+Au collisions at several different collision energies are summarized in Figs.~\ref{fig2}, \ref{fig3}, \ref{fig4} and \ref{fig5}.

%
% Fig. 2
%
The extracted values of v$^{even}_{1}(\pT)$  for 0-10\%, 10-20\% and 20-30\% centrality selections are shown  in Fig.~\ref{fig2}; the solid line in panel (a) indicates the hydrodynamic calculations~\cite{Retinskaya:2012ky}, that in good agreement with our measurements, the inset displays the results of the associated momentum conservation coefficient, $C$, obtained for several centralities at $\sqrt{s_{_{NN}}} = 200$~GeV. The v$^{even}_{1}(\pT)$  values show the characteristic pattern of a change from negative  v$^{even}_{1}(\pT)$ at low $\pT$ to positive v$^{even}_{1}(\pT)$ for $\pT > 1$ ~GeV/c, with a crossing point that slowly shifts with $\sqrt{s_{_{NN}}}$. They also indicate that v$^{even}_{1}$ increases as the collisions become more peripheral, as might be expected from the centrality dependence of $\varepsilon_1$.

%
%%%%%%%%%%%%%%%%%%%%%%%%%%%%%%%%%%%%%%%%%%%%%%%
%%%  Fig-3
%%%%%%%%%%%%%%%%%%%%%%%%%%%%%%%%%%%%%%%%%%%%%%%
%

%%%%%%%%%%%%%%%%%%%%%%%%%%%%%%%%%%%%%%%%%%%%%%%
Figure~\ref{fig2} shows the $p_T$ dependence of the v$_{n}$($n \ge 2$) measurements for 0-40\% centrality selection for a representative set of beam energies. Fig.~\ref{fig2} shows the v$_{n}$ dependence on $p_{T}$ and the harmonic number, $n$, with similar trends for each beam energy. 

%
%%%%%%%%%%%%%%%%%%%%%%%%%%%%%%%%%%%%%%%%%%%%%%%
%%%  Fig-4
%%%%%%%%%%%%%%%%%%%%%%%%%%%%%%%%%%%%%%%%%%%%%%%

The centrality dependence of v$_{n}$($n \ge 2$) is indicated in Fig.~\ref{fig4} for the same representative set of beam energies. Our measurements indicate a soft centrality dependence for the higher-order flow harmonics, which all decrease with decreasing the $\sqrt{s_{\rm NN}}$. These $v_n$ patterns may be related to the dependence of the viscous effects in the created medium, which lead to attenuation of $v_n$ magnitude. 

%%%%%%%%%%%%%%%%%%%%%%%%%%%%%%%%%%%%%%%%%%%%%%%
%%%  Fig-5
%%%%%%%%%%%%%%%%%%%%%%%%%%%%%%%%%%%%%%%%%%%%%%%

Figure~\ref{fig5} gives the excitation functions for the $p_T$-integrated v$_{2}$, v$_{3}$ and v$_{4}$  for $0-40\%$ central Au+Au collisions. They indicate monotonic trend for v$_{n}$ with $\sqrt{s_{\rm NN}}$, as might be expected for a temperature increase with $\sqrt{s_{\rm NN}}$.

\section{Conclusion} 
In summary, we have presented a comprehensive set of STAR anisotropic flow measurements for Au+Au collisions at $\sqrt{s_{\rm NN}}$ $=$ 7.7-200~GeV. The measurements use the two-particle correlation method to obtain the Fourier coefficients, v$_{n}$($n > 1$), and the rapidity-even dipolar flow coefficient, v$^{even}_{1}$. The rapidity-even dipolar flow measurements indicate the characteristic patterns of an evolution from negative v$^{even}_{1}(\pT)$  for $\pT ~<~1$~GeV/c to positive v$^{even}_{1}(\pT)$ for $\pT ~>~ 1$~GeV/c, expected when initial-state geometric fluctuations act along with the hydrodynamic-like expansion to generate rapidity-even dipolar flow. The v$_{n}$($n > 1$) measurements show a rich set of dependences on harmonic number $n$, $p_T$ and centrality for several collision energies. This set of measurements may provide additional constraints to test different initial-state models and to aid accuracy extraction of the temperature dependence of the specific shear viscosity.

\section*{Acknowledgments}
%=======================================
This research is supported by the US Department of Energy under contract DE-FG02-94ER40865.

\bibliography{BES_v1}

\end{document}